\documentclass[a4paper,12pt]{article}
\usepackage{geometry}
\usepackage[utf8]{inputenc}
\usepackage[T1]{fontenc} %
\usepackage[swedish,american]{babel}

\usepackage[autostyle]{csquotes}

\usepackage[backend=bibtex,style=authoryear,backref=false,uniquename=false,dashed=false]{biblatex} %
\makeatletter

\newrobustcmd*{\parentexttrack}[1]{%
  \begingroup
  \blx@blxinit
  \blx@setsfcodes
  \blx@bibopenparen#1\blx@bibcloseparen
  \endgroup}

\AtEveryCite{%
  }

\makeatother

\let\cite\parencite

\usepackage{url}
\def\UrlFont{\tt} %

\usepackage{hyperref}

\usepackage[svgnames]{xcolor}
\hypersetup{
	colorlinks,
	breaklinks,
	linkcolor=black,
	urlcolor=MidnightBlue,
	citecolor=MidnightBlue
}

\usepackage{color}
\usepackage{soul}
\usepackage{adjustbox} %
\usepackage{relsize} %
\usepackage{amssymb} %
\usepackage{amsthm}
\usepackage{xfrac}
\theoremstyle{definition}

\usepackage{paralist}
\usepackage{rotating} %
\usepackage{pifont}
\usepackage{wrapfig}
\usepackage{float}
\usepackage{mdframed}
\usepackage{footnote}

\usepackage{afterpage}
\usepackage{tikz}
\usepackage{tabularx}
\usepackage{enumitem}
\usepackage{multicol,caption}
\usepackage{lipsum}

\newenvironment{Figure}
  {\par\medskip\noindent\minipage{\linewidth}}
  {\endminipage\par\medskip}

\usepackage{comment}

\excludecomment{inlinefigures}
\includecomment{figuresinclude}
\usepackage[shadow,textsize=small]{todonotes}

\usepackage{xspace}
\newcommand{\ie}{\textit{i.e.,}\xspace}
\newcommand{\eg}{\textit{e.g.,}\xspace}

\newcommand{\incmt}[1]{\ignorespaces}

\newcommand{\sectioncaps}[1]{ \section{\textsc{#1}} } %

\begin{document}

\title{IPSME- Idempotent Publish/Subscribe Messaging Environment \texttt{\Large[v1.4x-18-g036cd77-0423+2023]} }
\author{
	Kim Nevelsteen \\
	\texttt{kim.nevelsteen@mitm.se} \\
	MiTM AB 
	\and
	Martin Wehlou\\
	\texttt{martin.wehlou@mitm.se} \\
	MiTM AB
}
\date{}
\maketitle

\begin{abstract}

The integration (interoperability) of highly disparate systems is an open topic of research in many domains. 
A common approach for getting two highly disparate systems to be interoperable, is through an agreed-upon protocol (\eg via standardization) or by employing a common framework.
The problem of integrating systems arises when many of these protocols/frameworks come into existence. 
Both, agreeing on protocols/frameworks and creating mappings between protocols takes time and effort. An interoperability solution must be scalable and should not require stakeholders to adapt to major changes in their system \ie systems should not need to be re-engineered as other systems are added, removed or replaced in the integration.

IPSME is introduced as a solution for integrating highly disparate systems. 
IPSME decouples the dependencies between interacting participants. 
Interoperability is achieved through dynamic translations, avoiding the need for agreed-upon protocols or frameworks. 
Scalability is achieved by not having a limitation on the number of messaging environments or the topological organization thereof.
IPSME is minimally invasive and through a network effect reduces the overall complexity of integrating many systems to linear. 
IPSME has been evaluated and thus far been tested in three use cases.

\end{abstract}

\phantom{.}

\noindent \textbf{keywords:} 
publish/subscribe; integration; interoperability; system of systems; mapping; scalability; evolution; architecture; metaverse; heterogeneous; systems; internet of things; virtual; idempotence; invocation.

\begin{itemize}[label={},itemindent=-2em,leftmargin=2em]
\item Nevelsteen, Kim and Martin Wehlou (2021). 
``IPSME- Idempotent Publish/Subscribe Messaging Environment.''
In: \textit{International Workshop on Immersive Mixed and Virtual Environment Systems Proceedings (MMVE '21), Sept.28-Oct.1, 2021, Istanbul, Turkey}. ACM. \\
\textsc{doi}: \url{https://doi.org/10.1145/3458307.3460966}.
\end{itemize}

\sectioncaps{Introduction}

The integration (interoperability\nocite{bass2013-inpractice}) of highly disparate systems is an open topic of research in many domains, including Healthcare~\cite{barthell2004-integration,landman2011-phit}, Computer Video Games~\cite{morgan2009-simgaming}, Internet of Things~\cite{noura2019-interop, nevelsteen2016-delegation}, 
Pervasive Applications~\cite{nevelsteen2016-thesis}, and the Metaverse~\cite{dionisio2013-metaverse}. And, by extension Cyber Physical Systems~\cite{gurdur2018-cpsinterop}, which incorporates Internet of Things and the Metaverse~\cite{rehm2015-mediator}. 
The domain of this article can be generalized as the integration of virtual environments of disparate systems, potentially forming a `system of systems'~\cite{madni2014-sosi}.

A common approach for getting two highly disparate systems to be interoperable and communicate, is by having them speak the same protocol. The problem of integrating systems arises when many of these protocols/frameworks come into existence. If $n$ distinct systems are to be integrated, then the eventual complexity will be $n(n-1)/2$~\cite{noura2019-interop}.

To address the problem, this article provides -- not a call for standardization or a framework as dependency, which are too restrictive -- but, a set of conventions which enables integration by: `decoupling'~\cite{bass2013-inpractice} dependencies and incorporating `interest management'~\cite{nevelsteen2016-delegation},
allowing for mappings between protocols dynamically, specifying the integrations to be external to the systems being integrated, and the division of communicating systems into 'regions'~\cite{nevelsteen2016-delegation,nevelsteen2017-virtualworlddef}.
To provide interoperability and decouple dependencies, a publish/subscribe system (pubsub) is used in the set of conventions.
\citeauthor{wang2002-secureinternetscale}~\cite{wang2002-secureinternetscale} state that pubsub is ``a communication infrastructure that enables data access and sharing over disparate systems and among inconsistent data models''.
Interest management for each system is incorporated by having participants simply drop messages not understood or found interesting. 
Mappings between protocols can bridge several `layers of interoperability'~\cite{noura2019-interop, halevy2005-datamix}, and in the set of conventions, mappings are specified in a manner so as to resolve schema heterogeneity through `decomposition'~\cite{selberg2008-evosos} of the problem.
And, the dividing of systems into regions provides the required `scalability'~\cite{bass2013-inpractice} for the system~\cite{noura2019-interop,nevelsteen2016-delegation}.
The set of conventions together form what is introduced in this article as an
Idempotent Publish/Subscribe Messaging Environment (IPSME).

To the best of the authors knowledge, as of this writing, IPSME is the only interoperability solution that is, independent of the systems being integrated, and allows for the integration of $n$ systems with linear complexity, while maintaining scalability.

\sectioncaps{Problem Statement}

A common approach for getting two highly disparate systems to be syntactically and semantically interoperable (assuming the systems can communicate \ie technical interoperability exists~\cite{noura2019-interop}), is by having them speak the same protocol, either: 
by having a pre-agreed upon protocol with possibly varying implementations on each system, or by employing a framework (\eg Software Development Kit), specifically implemented for each system, to ensure communication. Integration is particularly a problem when dealing with existing `legacy' systems~\cite{madni2014-sosi}, which speak different protocols (or divergent versions), that are difficult to alter.

The problem of integrating systems arises when many of these protocols/frameworks come into existence. Agreeing on protocols takes time and doesn't necessarily satisfy all `stakeholders'~\cite{madni2014-sosi}. This delay can stifle fast paced innovation~\cite{shapiro1999-standards}.  
Different protocols can be bridged by making a mapping between the several layers of interoperability.
A key issue with mappings is that they also take a considerable amount of effort \eg ``in a typical data integration scenario, more than half of the effort (and sometimes up to 80 percent) is spent on creating the mappings, and the process is labor-intensive and error-prone''~\cite{halevy2005-datamix}.
Mappings must not only satisfy syntactical and semantic interoperability, but also the conceptual layers of interoperability~\cite{noura2019-interop}.
There is not necessarily a single correct mapping~\cite{halevy2005-datamix} and when protocols are updated, any existing mappings must also be updated, exaggerating the problem.

\sectioncaps{Related Work} %

When faced with the problem of integrating highly disparate systems, many~\cite{landman2011-phit,morgan2009-simgaming,branton2011-interop,barthell2004-integration} argue that interoperability can be achieved through standardization, but standardization only has limited success~\cite{halevy2005-datamix}.
Features are either lacking at the time of standardization or the standard can become bloated with features requested from various stakeholders. 
It is impossible for a standardization committee to foresee all possible usages of a given standard.
A considerable amount of time is required to create and update a standard~\cite{noura2019-interop}, 
\eg the proposed standard being outdated before it is standardized.

In Internet of Things, where there is currently active research in interoperability, gateways employing `mediators' have been developed between devices to ``bridge between different specifications, data, standards, and middleware's etc.''~\cite{noura2019-interop} These gateways can be expandable via plugins to upgrade interoperability with more systems, but ``this approach has [a] limitation on scalability''~\cite{noura2019-interop}. Open challenges for interoperability include scalability and 
that stakeholders should not have to ``adapt to major changes in their system; the solution should not be dependent on their system''~\cite{noura2019-interop}.

\sectioncaps{Introducing IPSME} %

IPSME is introduced to enable the integration of highly disparate systems. 
Rather than a framework or standardized protocol, IPSME is a set of conventions that forms an `architecture'~\cite{bass2013-inpractice} where any participant to talk to any other participant, without need for a central authority and without standardization, provided groups of participants speak the same protocol.

\phantom{.}

\noindent IPSME defines the following conventions:
\begin{itemize}[nosep, leftmargin=4ex, labelwidth=*]
\item A messaging environment (ME) is defined as:
	\begin{itemize}
	\item A pubsub system which receives messages and relays those messages to all subscribed participants;
	\item Messages must be idempotent or identifiable as duplicates.
	\end{itemize}	
\item Each participant:
	\begin{itemize}
	\item sends and receives messages in a (local) ME;
	\item simply ignores messages, if not understood.
	\end{itemize}	
\item Translation of messages is done by having a participant listen to messages on a (local) ME and sending out translated messages. 
\item Communication across ME boundaries is through a \textit{reflector} pair: a participant listener with a counterpart in another ME, which resends messages there. Communication between reflectors is left unspecified and completely up to the author(s) of the reflectors, as is the selection of messages to resend.
\end{itemize}

\phantom{.}

\noindent 
The author(s) of a participant is free to implement their own message format as long as the above conventions are met. The set of conventions is purposely kept non-restrictive for easy adoption. 

\subsection{Interoperability}

A local ME employs the usage of a readily available pubsub resource. On the various operating systems, there is usually a platform-specific messaging system for inter-process communication (IPC) where pubsub can be used \eg NSNotificationCenter on \mbox{macOS}/iOS and MSMQ on Windows. 
If a platform-specific messaging system is not available, any other available pubsub can be used, as long as all participants that are to be local to that ME, know how to access the ME. 
If more than one pubsub is utilized on a platform and they are to interact, they must be interconnected by a pair of reflectors. 
Participants of a ME are usually processes running on the platform.
Participants can be both publishers and/or subscribers in a ME. The pubsub in a ME serves no other purpose than to relay published messages by broadcasting them to subscribers.

Because pubsub is a broadcast system, messages in IPSME are required to be `idempotent`~\cite{brown2002-idempotent}, so as to promote asynchronous communication, by reducing the amount of acknowledges required, and that identifiable duplicates can be eliminated.
If messages are passed across multiple MEs a universally unique identifier (\eg \texttt{UUID} or \texttt{GUID}) can be employed to help achieve idempotence.
Idempotent messages can be processed multiple times by a participant, but the processing of each message must give the same result after the application of the initial message. 
Because messages are broadcast using pubsub, messages are the `implicit invocation`~\cite{garlan1993-introarchitecture} of potentially multiple participants.
This does add more complexity to the sender, since the sender must handle zero or multiple responses \eg pick the best response or reply to all the responses within a given timeframe.

Rather than trying to obtain interoperability by enforcing a predefined structure or data model in messages~\cite{eugster2003-pubsub}, 
IPSME does not specify a format for message content \ie it is possible to use strings, binary or any of the topic-, content- or type-based \mbox{pubsub} schemes~\cite{eugster2003-pubsub}. 
By not specifying message content IPSME avoids having a predetermined `expressiveness'~\cite{carzaniga2001-designwide}, perhaps having many simultaneously.
Participants send messages in their own protocol and only participants that understand those messages will be able to process them \ie partitioning the semantic and syntactic space into any number of separate spaces. 
One of the main tenets that enables interoperability for IPSME is the interest management of each participant; if a participant does not understand a message, it simply drops the message and continues processing. This can lead to scenarios where certain participants might want affirmation that another participant has received the message. It is possible to send return messages (\eg Remote Procedure Calls or Acknowledges) through IPSME, but such a return is not defined in the IPSME specification. 
The IPSME conventions are minimally invasive for existing systems, since those systems can continue to speak the protocol that was previously implemented, but can be externally bound to a ME.

Any authentication or message security~\cite{uzunov2016-securitysurvey} is up to the author(s) of the participants \ie security is left as peripheral to this discussion, but has been taken into consideration during the design of IPSME.
If standard and/or central authentication services are required, such a service can be provided through the use of a translator.

\subsection{Mappings}

If each participant sends messages in their own protocol, communication is limited to the number of participants that understand the sent messages. To broaden the set of participants that understand a message, specialized participants that translate messages can be inserted in to a ME. These translators listen for messages that adhere to a certain protocol, translate them to a different protocol (\ie a mapping) and send out the translation; translating participants are considered mediators between other participants. 

Between all participants of a local ME and through to other MEs via reflectors, translations have a \textit{network effect}. Translations are transitively applicable to other participants \eg if a translation $X$ translates from participant $A$ to $B$, then any participant $C$, that can communicate with $A$, can also communicate with $B$, via $X$.
The network effect of translations has the potential to reduce the complexity of integrating $n$ participant nodes from an exponential $n(n-1)/2$, to a linear $(n-1)$.

The use of translators in this manner means IPSME alleviates the problem of resolving schema heterogeneity through the decomposition of the problem.
The translators can collectively divide the problem vertically, horizontally~\cite{uzunov2016-securitysurvey} or even incrementally. %
A major advantage of this architecture is a human related one. Authors of participants have a very limited scope of other communicating participants they must take into account. 

Participant communication is not constrained to be via a ME. Participants can negotiate to communicate directly allowing for `explicit invocation'~\cite{garlan1993-introarchitecture}. It is the responsibility of the participants to negotiate such a communication, and beyond the IPSME specification.

\subsection{Scalability}

Through the use of pubsub, IPSME decouples the production and consumption of messages, increasing scalability by decoupling dependencies (\ie time, space or synchronization) between interacting participants. 
IPSME is not fixed to specific properties such as expressiveness~\cite{carzaniga2001-designwide} 
or those related to quality of service~\cite{eugster2003-pubsub}, that can affect scalability.
A single ME can take advantage of being centralized, but (for the integrated systems as a whole) a centralized architecture or a hierarchical topology should be avoided so as to promote scalability; a centralized architecture being a bottleneck and single point of failure, and a hierarchical topology having possible performance problems~\cite{carzaniga2001-designwide}.

Expecting all prospective participants to be connected to the same ME is impractical; not all prospective participants would easily route to a single ME and a single ME would certainly be overloaded. Participants can be divided (\ie into regions) using multiple MEs. IPSME specifies participants should connect to a local ME, but places no limitation on the number of MEs or the `topology organization'~\cite{uzunov2016-securitysurvey} thereof. IPSME specifies reflectors for communication across ME boundaries, and it is this communication that allows MEs to be connected and organized into a general graph topology. 

A reflector is a participant in a ME that listens and filters for particular messages that should be routed to another ME. The reflector communicates directly with a reflector (\ie its counterpart) in that other ME. Upon receiving a message, the counterpart publishes the message to its local ME; participants of that local ME will receive messages from the distant ME transparently \ie without being aware of any communication complexity thereof. 
Communication between two reflectors is left as undefined and is completely up to the author(s) of those participants. 
A reply message is routed back through reflectors in a reverse fashion. Similar to how adding translators adds functionality without changing the existing system, reflectors can also be inserted into a ME without changing existing implementations; to the participants using a reflector the ME is simply expanded.

\begin{inlinefigures}

\setcounter{figure}{0}
\renewcommand{\figurename}{Sequence Diagram}

\begin{figure}[!ht]
\includegraphics[width=1\columnwidth]{imgs/wsd-MinecraftDoom_integration-IPSME-short.png}
\vspace*{-6mm}
\caption[]{\label{wsd:minecraftdoom} Minecraft/Doom integration: messaging for teleporting between Minecraft \& Doom3, and back} %
\end{figure}

\end{inlinefigures}

\begin{figuresinclude}

\newpage

\begin{multicols}{2}

\setcounter{figure}{0}
\renewcommand{\figurename}{\scriptsize Sequence Diagram}

\begin{Figure}
\includegraphics[width=\textwidth,height=\textheight,keepaspectratio]{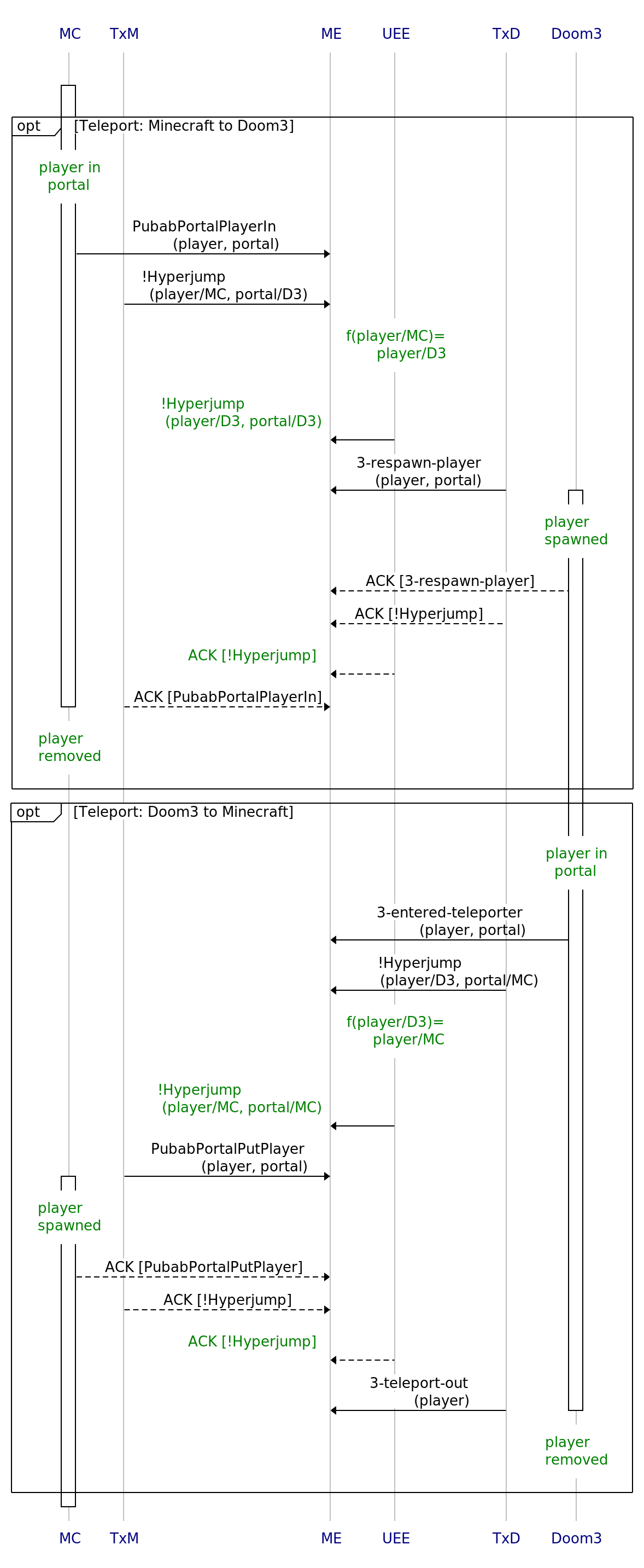}
\vspace*{-8mm}
\captionof{figure}{\label{wsd:minecraftdoom}\scriptsize Minecraft/Doom integration: messaging for teleporting between Minecraft \& Doom3, and back} 
\end{Figure}

\setcounter{figure}{1}
\renewcommand{\figurename}{\scriptsize Figure}

\begin{Figure}
\begin{tabularx}{225pt} { %
  >{\centering\arraybackslash}X %
  >{\centering\arraybackslash}X %
}
\begin{tikzpicture}[node distance={15mm}, main/.style = {draw, circle}] 
\node[main] (1) {$\texttt{MC}$}; 
\node[main] (3) [below right of=1] {$\texttt{UEE}$}; 
\node[main] (4) [above right of=3] {$\texttt{Doom3}$}; 
\draw (1) -- node[below left]{TxM} (3); 
\draw[dotted] (1) -- (4); 
\draw (3) -- node[below right]{TxD} (4); 
\end{tikzpicture} 
& 
\begin{tikzpicture}[node distance={15mm}, main/.style = {draw, circle}] 
\node[main] (1) {$\texttt{MC}$}; 
\node[main] (2) [above right of=1] {$\texttt{MOO}$}; 
\node[main] (3) [below right of=1] {$\texttt{UEE}$}; 
\node[main] (4) [above right of=3] {$\texttt{Doom3}$}; 
\draw (1) -- node[below left]{TxM} (3); 
\draw[dotted] (1) -- (4); 
\draw (3) -- node[below right]{TxD} (4); 
\draw (1) -- (2); 
\draw (2)[dotted] -- (3); 
\draw[dotted] (2) -- (4); 
\end{tikzpicture} 
\end{tabularx}
\vspace*{-2mm}
\captionof{figure}{\label{fig:participantgraphs}\scriptsize  Translation graphs for three and four participants, respectively 
(solid lines are implemented translation; dotted lines are gained through a network effect).}
\end{Figure}

\setcounter{figure}{1}
\renewcommand{\figurename}{\scriptsize Sequence Diagram}

\begin{Figure}
\includegraphics[width=\textwidth,height=\textheight,keepaspectratio]{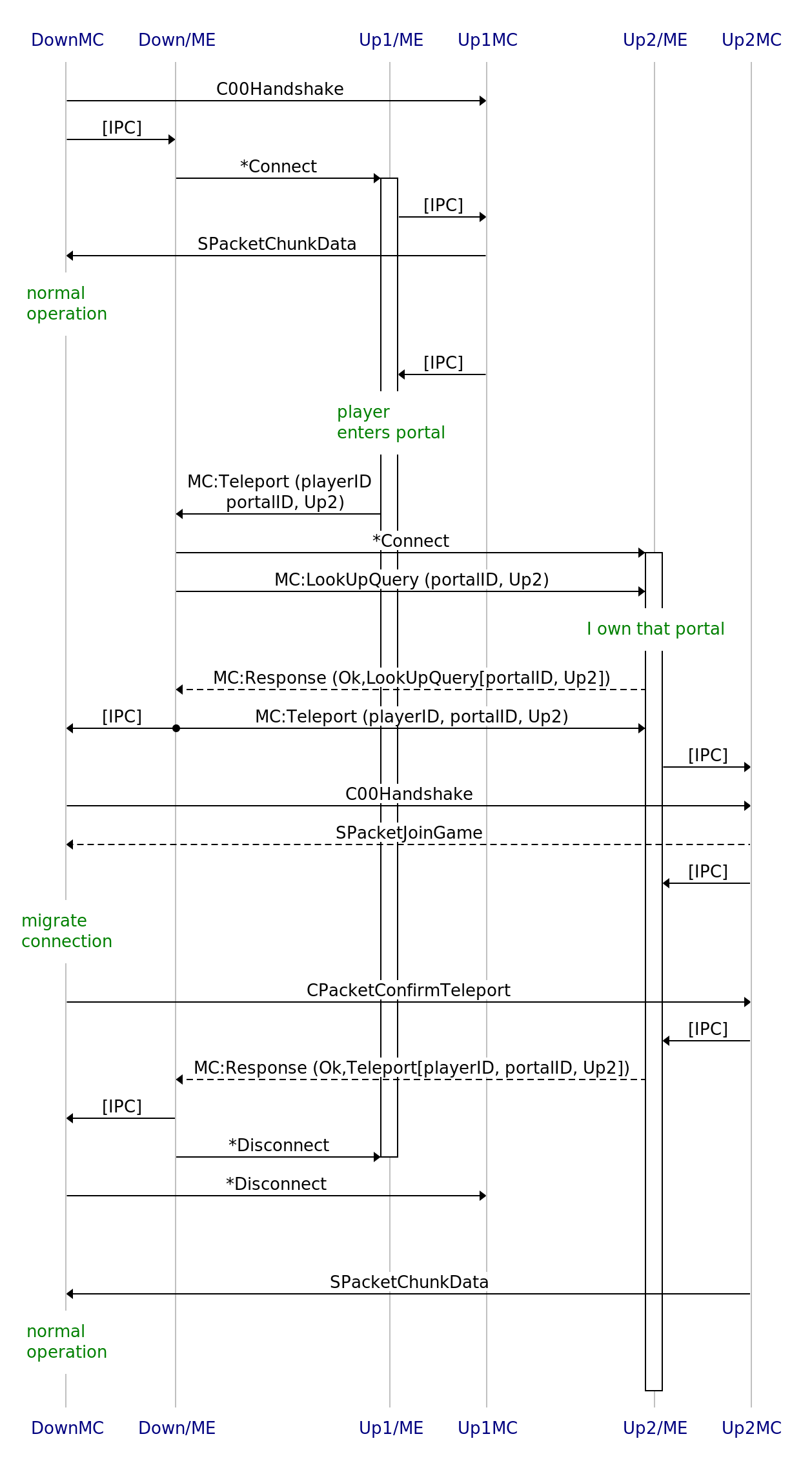}
\vspace*{-8mm}
\captionof{figure}{\label{wsd:minecraftmetaverse}\scriptsize Minecraft-Metaverse via MiM: an illustration of using direct communication together with IPSME}
\end{Figure}

\end{multicols}

\newpage

\end{figuresinclude}

\sectioncaps{Evaluation}

IPSME has thus far been tested in the following three use cases: 
\begin{inparaenum}[]
\item the \textit{Minecraft/Doom integration}\footnote{ \textit{Minecraft/Doom integration forming a Metaverse}: \url{https://youtu.be/knKZd15rhJE} },
\item \textit{Medical Resource Scheduling}\footnote{ \textit{Medical Resource Scheduling}: \url{https://youtu.be/m_b1PotByx0} }, and
\item a \textit{Minecraft-Metaverse via MiM}\footnote{ \textit{Insane Minecraft Teleportation, in VR, via MiM}: \url{https://youtu.be/ORWto-Oo1W4} }.
\end{inparaenum}

A proof-of-concept dubbed \textit{Metaverse prototype}\footnote{ \textit{Metaverse prototype verbose demo}: \url{https://youtu.be/etKJMUyPn-8} } was a \mbox{precursor} to the three use cases, integrating the video games Minecraft and Doom3, and the virtual world of LambdaMOO. The proof-of-concept did not use IPSME; the integration between Minecraft and Doom, and the integration between Minecraft and LambdaMOO were implemented directly in each participant. No integration from Doom to LambdaMOO was achieved.

\subsection{Minecraft/Doom integration}

In the Minecraft/Doom integration use case (depicted in Sequence Diagram~\ref{wsd:minecraftdoom}), both Doom and Minecraft were slightly altered to: connect to the local ME, and expose an API for teleporting in and out of their respective virtual environments. %
A single local ME was used, with the following components attached to it: Doom3, Minecraft~(MC), a trivial universal interface component (UEE) and a Minecraft translation component (TxM).
The Doom3 translation participant (TxD) was implemented in the UEE component.
In hindsight, it would have been sufficient for the translation components (TxM and TxD) to call the Doom3 and MC APIs directly, reducing the required changes to Doom3 and Minecraft.
Although all components were on the same operating system (\ie macOS), communication spanned two programming languages: Java for MC and TxM, and C\verb!++! for the ME, UEE and Doom3. 
TxM was responsible for translating the custom Minecraft protocol to a custom Hyperjump protocol \ie syntactical mapping. UEE understood the Hyperjump protocol and was also responsible for translating to the custom Doom3 protocol. 
It should be noted that each signal in Sequence Diagram~\ref{wsd:minecraftdoom} is a broadcast to all other components through the local ME. A signal is a reply to the signal that precedes it, even though no arrow is depicted showing the arrival of the broadcast.  

Minecraft was altered so that when a player enters a portal, a \texttt{PubabPortalPlayerIn} message is broadcast, which is understood by TxM causing it to subsequently send out a \texttt{Hyperjump} message. UEE accepts the \texttt{Hyperjump} message and sends out a translated message, with the Minecraft player (\texttt{player/MC}) translated to the corresponding Doom3 player (\texttt{player/D3}), via internal semantic and context mapping of player profiles and inventories. 
TxD understands the \texttt{Hyperjump} message and translates the message to the Doom3, \texttt{3-respawn-player} message. 
It would have been sufficient for TxD to call Doom3 via an API call, but instead Doom3 accepts the \texttt{3-respawn-player} message via ME and spawns the player. An acknowledge is sent back through in reverse order, so that the player can be removed from the Minecraft virtual environment. A teleport from Doom3 to Minecraft is handled similarly. 

Even with such a small number of participants, the benefit of the network effect of translations is noticeable. To obtain the complete integration of three participants (\ie MC, UEE and Doom), a fully connected translation graph is required; see Figure~\ref{fig:participantgraphs}~(left).
The two translations provided by TxM and TxD were implemented, but a third translation from MC to Doom3 is achieved through the network effect of transitively applying TxM and TxD. The complexity of the full integration is thereby reduced from $3(3-1)/2=3$ to $(3-1)=2$.

In the precursor proof-of-concept, LambdaMOO was integrated with Minecraft. 
If LambdaMOO were to be integrated into the Minecraft/Doom integration, using IPSME instead, a MOO participant would be added to the translation graph, with an implemented translation between MOO and MC; see Figure~\ref{fig:participantgraphs}~(right). The integration of MOO and UEE, and also MOO and Doom3, would be obtained through the network effect.
The complexity of the four fully integrated participants is $(4-1)=3$, rather than $4(4-1)/2=6$.

\begin{inlinefigures}

\setcounter{figure}{0}
\renewcommand{\figurename}{Figure}

\begin{figure}
\vspace*{6mm}
\begin{tabularx}{225pt} { %
  >{\centering\arraybackslash}X %
  >{\centering\arraybackslash}X %
}
\begin{tikzpicture}[node distance={15mm}, main/.style = {draw, circle, scale=0.9}] 
\node[main] (1) {$\texttt{MC}$}; 
\node[main] (3) [below right of=1] {$\texttt{UEE}$}; 
\node[main] (4) [above right of=3] {$\texttt{Doom3}$}; 
\draw (1) -- node[below left]{TxM} (3); 
\draw[dotted] (1) -- (4); 
\draw (3) -- node[below right]{TxD} (4); 
\end{tikzpicture} 
& 
\begin{tikzpicture}[node distance={15mm}, main/.style = {draw, circle, scale=0.9}] 
\node[main] (1) {$\texttt{MC}$}; 
\node[main] (2) [above right of=1] {$\texttt{MOO}$}; 
\node[main] (3) [below right of=1] {$\texttt{UEE}$}; 
\node[main] (4) [above right of=3] {$\texttt{Doom3}$}; 
\draw (1) -- node[below left]{TxM} (3); 
\draw[dotted] (1) -- (4); 
\draw (3) -- node[below right]{TxD} (4); 
\draw (1) -- (2); 
\draw (2)[dotted] -- (3); 
\draw[dotted] (2) -- (4); 
\end{tikzpicture} 
\end{tabularx}
\vspace*{-2mm}
\caption[]{\label{fig:participantgraphs} {Translation graphs for three and four participants, respectively 
(solid lines are implemented translations; dotted lines are gained through a network effect).} }
\end{figure}

\end{inlinefigures}

\subsection{Medical Resource Scheduling}

The primary focus of the Medical Resource Scheduling use case was the ability of various Care Planner components to dynamically find and negotiate with various medical resource providers (\eg operating rooms). Although not in the use case, the system was designed such that each of the various components could have been on different platform. The entire system is design to be scalable and allow various stakeholders to negotiate without the need for a central authority. A large degree of fault tolerance was built into the use case, by detecting failed participants and executing multiple copies of the same participant to achieve redundancy.

\begin{inlinefigures}

\setcounter{figure}{1}
\renewcommand{\figurename}{Sequence Diagram}

\begin{figure}[!t]
\includegraphics[width=1\columnwidth]{imgs/wsd-MinecraftMetaverse-IPSME-short.png}
\vspace*{-6mm}
\caption[]{\label{wsd:minecraftmetaverse} Minecraft-Metaverse via MiM: an illustration of using direct communication together with IPSME  }
\end{figure}

\end{inlinefigures}

\subsection{Minecraft-Metaverse via MiM}

The most recent use case, a Minecraft-Metaverse via MiM (depicted in Sequence Diagram~\ref{wsd:minecraftmetaverse}), was an attempt to link existing Minecraft server instances, without altering the Minecraft client or server; this meant manipulating the streaming data of the proprietary Minecraft network protocol (a task handled by components with a -MC \mbox{suffix}). 
The normal stream between client and server was redirected to be: from client (not depicted in Sequence Diagram~\ref{wsd:minecraftmetaverse}), through the downstream -MC component, DownMC, through one of the upstream -MC components, Up1MC or Up2MC, to a corresponding Minecraft server (not depicted). 
Every -MC component had a corresponding local ME \ie the downstream participant, Down, and upstream participants, Up1 \& Up2, connected to their corresponding local MEs on the client-side and server-side, respectively. 
Reflectors dynamically connected downstream participants to any number of upstream participants, forming a star topology for a single Minecraft client.
The Protocol Buffers\footnote{Protocol Buffers: \url{https://developers.google.com/protocol-buffers} } interface description language was used to generate a protocol between upstream and downstream participants.
Events (detected in the Minecraft protocol stream or received through the MEs) were shared between Up/Down and -MC component pairs (\eg Up1 and Up1MC) via IPC.
The programming language for all components was Java, but components resided on various operating systems \eg macOS, Linux and Windows.  

Sequence Diagram~\ref{wsd:minecraftmetaverse} depicts how, after receiving a \texttt{MC:Teleport} message from Up1, the direct (proprietary Minecraft protocol) communication between DownMC and Up1MC is migrated over to be between DownMC and Up2MC instead. 
The sequence begins with a successful connection to Up1MC and Up1 \ie the packet \texttt{SPacketChunkData} being sent from server to client, in normal operation. 
Up1MC detects the movement of the player in the Minecraft protocol and shares that with Up1 via IPC. 
Since Minecraft servers were not altered, portal lists were kept in upstream components. 
Up1 determines that if a player has entered a portal, broadcasting out a \texttt{MC:Teleport} message in that event, which is then passed over to Down/ME via reflectors. Down reads the destination `Up2' out of \texttt{MC:Teleport} and sets up a connection to the potential destination; a \texttt{MC:LookUpQuery} is broadcast out by Down, which is reflected to all connected upstream participants. If no valid response to \texttt{MC:LookUpQuery} is received within reasonable time, \texttt{MC:Teleport} is dropped by Down and normal operation continues. As depicted, Up2 receives \texttt{MC:LookUpQuery}, confirms that it owns the exiting end of the portal and broadcasts out a corresponding \texttt{MC:Response}. 
Reflectors carry the response back to Down triggering: 
a rebroadcasting of the \texttt{MC:Teleport} message, notifying Up2 of the players arrival; and,
an IPC call to DownMC to pause the communication with Up1MC and open direct communication with Up2MC.
When DownMC detects a successful connection with Up2MC (\ie \texttt{SPacketJoinGame} received), the active connection is migrated from Up1MC to Up2MC. 
A player teleport is simulated originating from the Minecraft server (\ie inserted in the Minecraft protocol stream, not depicted), and confirmed with a \texttt{CPacketConfirmTeleport} reply from the client. Up2MC notifies Up2 (via IPC), when the simulated player teleport is complete. Up2 broadcasts out a corresponding \texttt{MC:Response}, which signals Down to disconnect from Up1, which was hosting the source end of the portal, and subsequently DownMC from Up1MC. 

\subsection{Discussion}

Rather than having to wait for features to be added to a standard, IPSME allows a translating participant to be added dynamically to the system, mediating communication with a mapping. 
Each translating participant only provides the required mapping, avoiding the problem of bloat. If many redundant translating participants are present, those which are not widely used can be detected, removed and replaced with a more minimal set of translators.
Since translating participants can be added dynamically, it is possible to avoid the delay of waiting for an agreed upon standard. 
The same dynamism ensures openness to unforeseeable future requirements.

IPSME can largely be considered a generalization (in software) of Internet of Things gateways as mediators, with the dynamic insertion of translating participants being similar to gateway plugins. The advantage of IPSME being that it offers a network effect by which translations become transitively applicable, removing the requirement that all  mappings must be one-to-one.
IPSME provides for scalability when the gateway approach was noted as being limited. Also, because of the dynamism of IPSME, stakeholders are not forced to adapt their systems to major changes in protocols; which solves that open challenge in the domain of Internet of Things.

\sectioncaps{Conclusion} %

In this article, IPSME is introduced as a solution for integrating highly disparate systems. 
Through the use of pubsub, IPSME decouples the production and consumption of messages, increasing scalability by removing dependencies (\ie time, space or synchronization) between interacting participants. 
Rather than requiring two systems to speak the same protocol, IPSME achieves interoperability through translations that can be dynamically added to the system, avoiding the need for pre-agreed upon protocols or frameworks, and also avoiding any delay that is incurred through coming to agreement. 
IPSME is minimally invasive and can be used to integrate legacy systems or even the most difficult interoperability cases~\cite{madni2014-sosi}.

Translations in IPSME can handle several layers of interoperability. 
The amount of effort it takes to create a mapping might not be reduced, but through the network effect granted by IPSME, translations can potentially be reused, reducing the exponential complexity of fully integrating many systems to linear.
And, when protocols are updated, rather than update each mapping, with IPSME a translation can be added that maps from the original protocol to the updated one and the integrated systems continue.

\label{section:future_work}

The intent is for IPSME to be broadly accepted for the integration of highly disparate systems, but there are environments which have specialized requirements \eg the medical field requires security. The set of conventions defined here as IPSME is only the primary layer of a multilayered system; additional layers (\eg to address service discovery and security) are left for subsequent publications.

\phantom{Adding cites that appear in final version}
~\nocite{kshemkalyani2008-distributedcomputing}
~\nocite{nakagawa2013-futuresos}
~\nocite{selberg2008-evosos}
~\nocite{wilenius2009-combinatorial}

\renewcommand*{\bibfont}{\small}
\renewcommand*{\UrlFont}{\ttfamily\smaller\relax}

\printbibliography[title=\textsc{References}]

\end{document}